\documentclass[pre,aps,twocolumn,showpacs]{revtex4}
\usepackage{epsfig}
\include{graphics}
\begin{document}

\title{Noise induces rare events in granular media}

\author{Evgeniy Khain$^1$ and Leonard M. Sander$^2$}
\affiliation{$^1$Department of Physics, Oakland University, Rochester, MI 48309, USA}
\affiliation{$^2$Department of Physics, University of Michigan, Ann Arbor, MI 48109-1120, USA}

\begin{abstract}
The granular Leidenfrost effect (B. Meerson et al, Phys. Rev. Lett. {\bf 91}, 024301 (2003), P. Eshuis et al, Phys. Rev. Lett. {\bf 95}, 258001 (2005)) is the levitation of a mass of granular matter when a wall below the grains is vibrated giving rise to a hot granular gas below the cluster. We find by simulation that for a range of parameters the system is bistable: the levitated cluster can occasionally break and give rise to two clusters and a hot granular gas above and below. We use techniques from the theory of rare events to compute the mean transition time for breaking to occur. This requires the introduction of a two-component reaction coordinate.

\end{abstract}


\maketitle

\section{Introduction}

Rare events \cite{Nicolis} in physical systems are important when they lead to dramatic changes in state. An example is an infrequent switch between two very different metastable states. Such noise induced transitions occur in physical, biological and chemical systems \cite{induced}; recent examples include optical \cite{optical}, cellular \cite{cellular}, and ecological \cite{ecological} systems. In this paper we treat such a transition in granular matter: the breaking of a levitated cluster (the ``granular Leidenfrost effect'') via a fluctuation.

In a generic rare event, a system spends a very long time wandering near one metastable state (these transitions are called typical events) and then rapidly jumps to another state due to a low-probability large fluctuation. Analyzing these switches in nonequilibrium systems is especially challenging since the rate to climb over the ``barrier" is not given by the classic Kramers formula,  $R \propto \exp(-\Delta E/kT)$ \cite{Gardiner}. In fact, such systems do not, in general,  have an energy landscape so that $\Delta E$ is not defined. Nevertheless, much of the phenomenology of equilibrium  transitions carries over to the non-equilibrium case. For example, there is often a point in the phase space of the system that looks like the classical saddle point or transition state, and  a most probable transition path, just as in Kramer's theory.
In systems with a few degrees of freedom progress can be made by writing down the master equation and employing the WKB method \cite{Kubo,DykmanPRE100,ElgartPRE70,Kessler,AM1}.

For systems with many degrees of freedom, such as the granular system we treat here (which has  thousands of interacting particles) simulation of the process is often the only choice. However, direct numerical simulation is challenging precisely because we are interested in improbable transitions. There is a large gap between the time scale of typical events and the rare events that interest us. To approach this problem a number of schemes have been formulated with the common idea of giving a higher weight to desired rare events. Notable among these are methods related to forward flux sampling \cite{Allen06,Allen06b,Adams10b} which are applicable to the non-equilibrium case. The general idea is to use short pieces of simulated trajectories to estimate the probability of making the transition.

Granular matter is a medium consisting of a large number of moving particles that collide inelastically \cite{review}. This simple feature, the inelasticity of collisions, implies that the system is intrinsically out of equilibrium, since some energy is lost in every collision. A system can be continuously driven by pumping in energy into it by either vibrating \cite{vibration} or organizing a shear \cite{shear}. If a balance is achieved, so that energy loss due to inelastic collisions is balanced by the external energy input, the system can reach a steady state.

One of the most intriguing steady states of this type is the  granular Leidenfrost state \cite{Leidenfrost1,Leidenfrost2}. The classical Leidenfrost effect occurs when a liquid droplet levitates over a hot surface \cite{Leidenfrostcl}. Due to rapid evaporation, the droplet hovers above hot vapor without falling down. A visually similar effect was found in granular matter when a large number of particles under gravity are driven from below by a vibrating ``thermal"  wall. In some parameter regimes, a solid  cluster of particles levitates above a hot granular gas \cite{Leidenfrost1,Leidenfrost2}. This  was experimentally observed by the groups of van der Meer and Lohse \cite{Leidenfrost2} in a system of vertically shaken grains. They were not only able to obtain the Leidenfrost state, but also observed that this state becomes unstable when the strength of driving exceeds a certain threshold \cite{Leidenfrost3}. The cluster becomes wavy and then breaks leading to convection. The phenomenon of thermal (Rayleigh-Benard like) convection in granular media was observed a few years earlier in molecular dynamics simulations \cite{convection1} and described theoretically in dilute granular gases \cite{convection2}. The transition from the Leidenfrost state to convection was further analyzed \cite{Leidenfrost4} experimentally, theoretically and in molecular dynamics simulations.

Here we report a different scenario which we observed in simulation (see also \cite{Isobe}). Once the temperature of the bottom ``thermal" wall, $T_b$, exceeds a certain threshold, the cluster breaks, but does not melt completely. The dynamics resembles a volcanic-like explosion, where the cluster creates a bump and then breaks; the hot gas fountains upward. This broken cluster persists for a very long time. We discovered that in a wide range of driving temperatures $T_b$, both states exist. This bistability brings about a possibility of a rare event: the cluster can break for low $T_b$ (well below the instability threshold) due to a large fluctuation. In the present paper we focus on this rare transition between the two states and compute the mean waiting time for rare cluster breaking.

One point of this exercise is to test the tools of rare-event simulation on an unfamiliar system. Previous work along these lines has mostly centered on the familiar case of nucleation of a cluster \cite{Adams10b,Valeriani,Allen08} for which there is a well-studied reaction coordinate, the cluster size. Here, as we will see, the reaction coordinate is less obvious, so that this is an instructive example.

\section{Granular Leidenfrost state and bistability}
Consider an ensemble of moving hard disks that collide inelastically; the particles of mass $m$ and diameter $d$ are placed into a two-dimensional system of width $L$ and height $H$; the bottom wall is thermal and maintained at temperature $T_b$; the gravity is in the negative $y$ direction. We employ periodic boundary conditions in the horizontal $x$ direction. The collisions are assumed to be binary and instantaneous; the inelasticity of collisions is modelled by the coefficient of normal restitution, $r<1$. Particles' velocities after the collision are related to their velocities before the collision; their tangential velocities remain unchanged, while the normal velocities after the collision are given by
\begin{equation}
 \left(
\begin{array}{c}
v^\prime_{i\parallel}  \\ v^\prime_{j\parallel}
\end{array} \right)=
\frac{1}{2}\left(
\begin{array}{cc}
1-r & 1+r
\\
1+r & 1-r
\end{array} \right)
\left(
\begin{array}{c}
v_{i\parallel}  \\ v_{j\parallel}
\end{array} \right)
\end{equation}
where the final velocities of the two colliding particles $i$ and $j$ are indicated by primes.

In molecular dynamics simulations, particles colliding with the bottom wall forget their previous normal velocities; the new normal velocities are taken from Maxwell-Boltzmann distribution with temperature $T_b$. Granular temperature decreases with height due to inelastic collision between the particles, as a result, in some parameter regime one can observe density inversion  \cite{Leidenfrost1}. The extreme case of this inversion is when the a solid cluster is levitating above the hot granular gas. Figure 1 (upper panel) shows that the cluster has an ordered crystalline structure \cite{Leidenfrost1}, similar to that observed in ``thermal" dense shear granular flows \cite{shear2,bistability2}. In the vicinity of a critical point, the theoretical progress can be made by deducing the van der Waals normal form \cite{normalform}.

\begin{figure}[ht]
\begin{center}
\includegraphics[width=3.0 in]{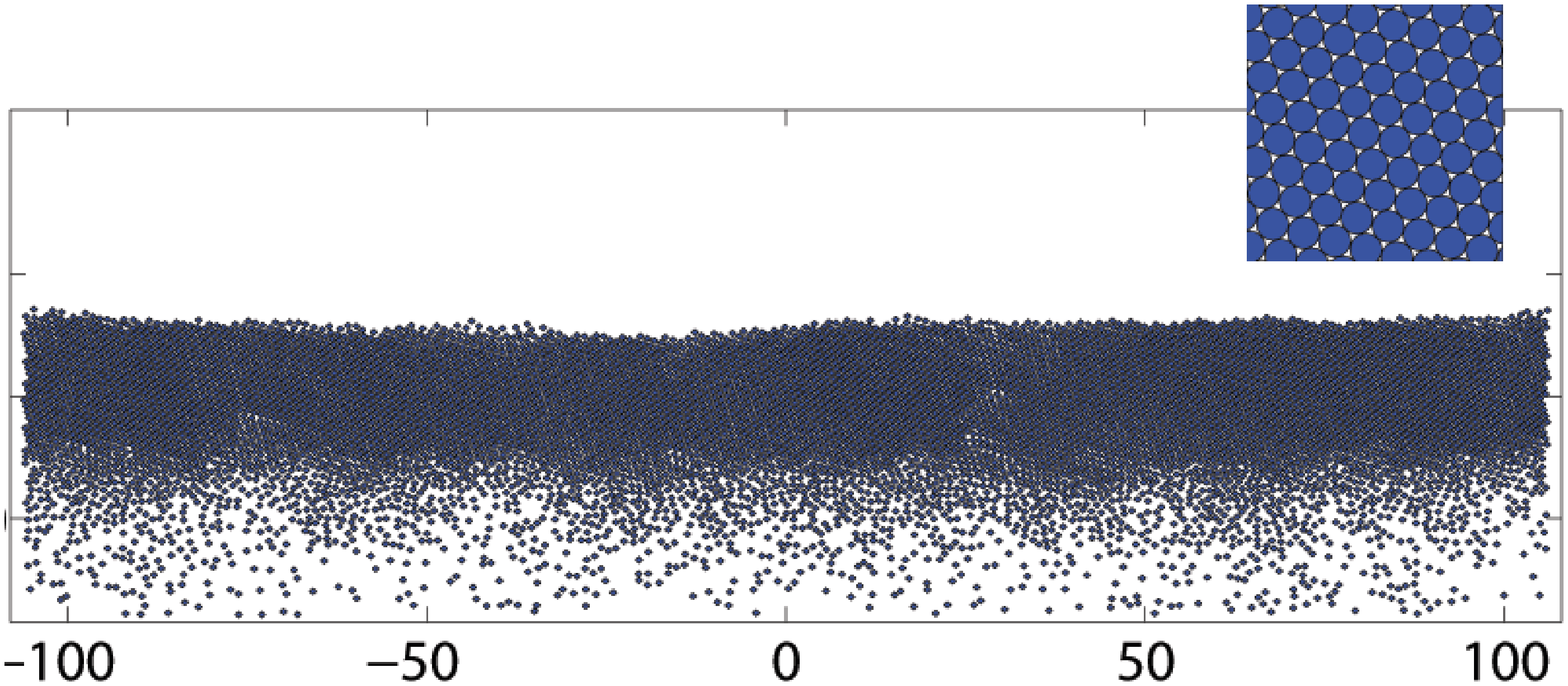}
\includegraphics[width=3.0 in]{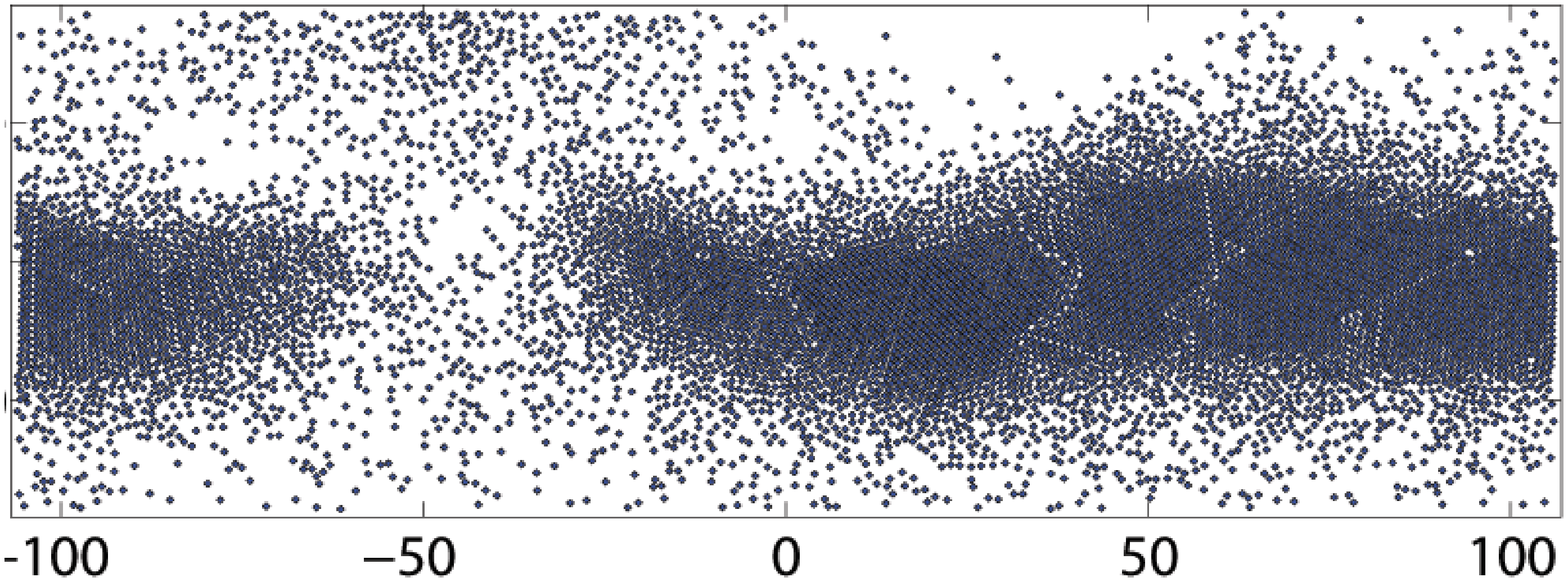}
\caption{Granular Leidenfrost state (upper panel). The broken cluster state (lower panel). Both states exist for the same parameters: $N=7500$, $r=0.928$, $T_b=1.46$, $m=d=1$.
\label{fig:cluster}
}
\end{center}
\end{figure}

When the temperature of the bottom wall exceeds a certain critical value, $T_{b, cr}$, the cluster breaks. This breaking resembles a volcanic explosion: one can see a fountain of hot particles streaming upward. This broken state seems to be stable (or very long lived); dense parts of a solid cluster coexist with hot gas and do not melt, see Figure 1 (lower panel). Interestingly, the broken state does exist also well below $T_{b, cr}$. Analyzing this phenomenon, we found bistability in a wide range of the temperatures of the bottom wall. Figure 2 shows the hysteresis, presenting the two states for various values of $T_b$. The states are characterized by the parameter $N_t$, the number of particles in a horizontal layer near the top wall. For the Leidenfrost state $N_t=0$ (the top wall can always be placed high enough); for the broken cluster state $N_t$ is high. We start following the system at low $T_b$, the cluster is quickly formed from the initially homogeneous system, so $N_t=0$. Then we slowly increase $T_b$ and the corresponding $N_t$ remains zero. At some value of $T_b$, the cluster breaks, $N_t$ jumps to a high value and continues increasing as $T_b$ increases. Then we can reverse the trend and slowly decrease $T_b$. $N_t$ slowly goes down and finally jumps to zero (the Leidenfrost state is formed), but this happens at a much smaller value of $T_b$, compared to the jump up.

\begin{figure}[ht]
\begin{center}
\includegraphics[width=3.0 in]{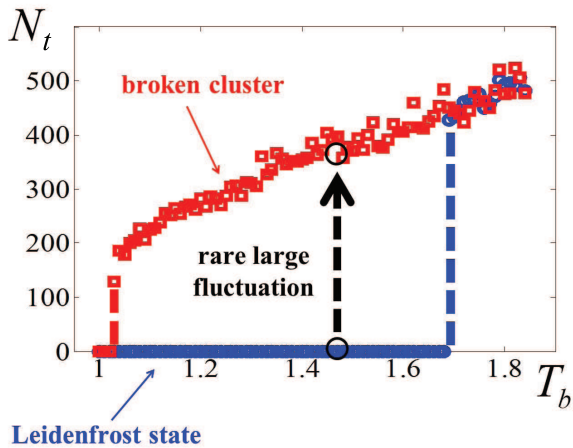}
\caption{Hysteresis. The number of particles in a layer near the top wall, $N_t$ as a function of the temperature of the bottom wall, $T_b$ for two different steady states. $N_t=0$ for the granular Leidenfrost state (unbroken cluster), $N_t>0$ for the broken cluster state, see Figure 1.
\label{fig:cluster}
}
\end{center}
\end{figure}

An intriguing phenomenon of bistability occurs also in other granular systems driven either by shear \cite{bistability2} or by a thermal wall. In the latter case, a granular cluster is formed along the wall opposite to the thermal wall. This cluster can exhibit spontaneous symmetry-breaking instability leading to phase separation: coexistence of dense and dilute regions of the granular media \cite{instability10}. In some region of parameters, both states are stable \cite{bistability3,bistability1}; spontaneous transitions between the two states were first observed by Argentina and coworkers \cite{bistability3}. Another system, which may exhibit bistability is a granular monolayer vibrated from below \cite{monolayer}. In particular, one can observe a transition to a state, where all the particles bounce vertically in phase with the vibrated bottom plate \cite{Soto}.

\section{Spontaneous cluster breaking: simulating rare events}

Let $T_b$ be inside the bistability range, away from the thresholds, and consider the granular Leidenfrost state. The cluster can still break (so, the system jumps to the broken cluster state, see the arrow in Figure 2) due to a rare large fluctuation. How can one compute the probability of such rare event? We briefly recall forward flux sampling (FFS) \cite{Allen06,Allen06b,Adams10b} for a simple one dimensional system. (In one dimension we always have an energy landscape, of course.)

Consider a particle in a double well potential.  The coordinates of the two wells are $X_A$ and $X_C$, and the coordinate of the potential maximum in between the two minima is $X_B$. The particle wanders around the minimum $X_A$, but a rare large fluctuation can bring it near $X_B$. Direct numerical simulations take a very long time (exponential in $\Delta E_{AB}$). In the FFS method we introduce ``barriers" between $X_A$ and $X_B$, see  Figure 3. The basic idea here is that we proliferate the system (put many identical copies)  on barrier $i$ and compute the fraction of these copies that go uphill to barrier $i+1$ before falling back to the well $X_A$; let us denote this number by $P_i$. Then the probability of moving from barrier $1$ to the peak $X_B$ can be written as a product of $P_i$'s. The total rate of the transition is obtained by dividing this product by $\tau_0$, the inverse frequency of crossing the barrier $X_1$ on the way up. The method speeds up the computation because we compute only small pieces of trajectory that go uphill. Because we make many copies, we enhance the sampling of the rare uphill trajectories.

\begin{figure}[ht]
\begin{center}
\includegraphics[width=3.0 in]{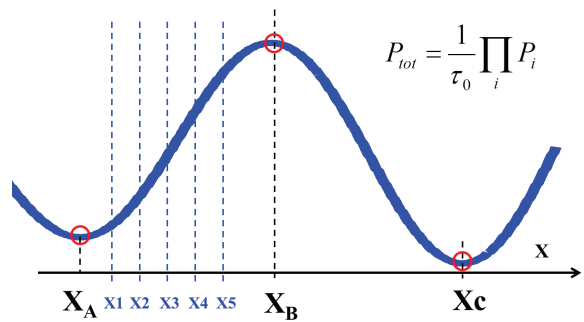}
\caption{Schematic representation of a barrier method for simulating rare events.
\label{fig:cluster}
}
\end{center}
\end{figure}

In our case, we have an extended system consisting of thousands of interacting particles. To proceed, we have to come up with a reduced dynamics in terms of a suitable reaction coordinate. Using the analogy of nucleation, we tried to find a one-dimensional reaction coordinate, for example, the maximum height of the particles. This led to rather ambiguous results. However, we found (by trial and error) that if we introduced a \emph{two component} reaction coordinate the description becomes rather simple.

We observed in simulation that when a cluster breaks, in a certain region the particles move up and to the sides, so the local vertical center of mass increases and the local number of particles goes down, see Figure 4. In order to quantify this, the system was divided into many vertical strips and in each strip two quantities were computed. The first one was the $y$-component of the center of mass of particles inside the strip, $y_{cm, i}$. The second was the number of particles in the strip, $n_i$.   Since we do not know where the cluster is going to break, the first reaction coordinate is taken to be the maximum of all $y_{cm, i}$ (minus the overall vertical center of mass of the system and scaled by the system height) and the second reaction coordinate is the minimum of all $n_i$ (scaled by the average number of particles in a stripe):
$$ \delta_1  = \max \left[\frac{1}{H} (Y_{cm,i} - \bar{Y}_{cm}) \right ], \quad \delta_2  = \min \left(\frac{N_i}{\bar{N}_{i}} \right ). $$

\begin{figure}[ht]
\begin{center}
\includegraphics[width=3.0 in]{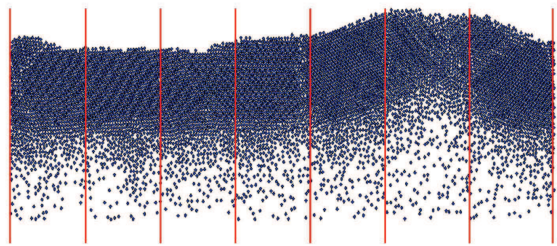}
\caption{Introducing two effective reaction coordinates for the reduced dynamics. In each vertical strip we compute the center of mass height and the number of particles. See text.
\label{fig:cluster}
}
\end{center}
\end{figure}

Figure 5 shows the phase plane of these two effective reaction coordinates. The granular Leidenfrost state is in the upper left corner of the diagram ($\delta_1$ is close to $0$, $\delta_2$ is close to $1$), the broken cluster state is in the lower right corner. The dynamics of the system is characterized by a trajectory on this phase diagram; each point represents the position of the system at a certain time. Figure 5 shows the typical time evolution of the system. The system spends an exponentially long time near the first ``well" (unbroken cluster state), then quickly ``climbs the hill" and goes to the second ``well", spending an exponentially long time there.

\begin{figure}[ht]
\begin{center}
\includegraphics[width=3.5 in]{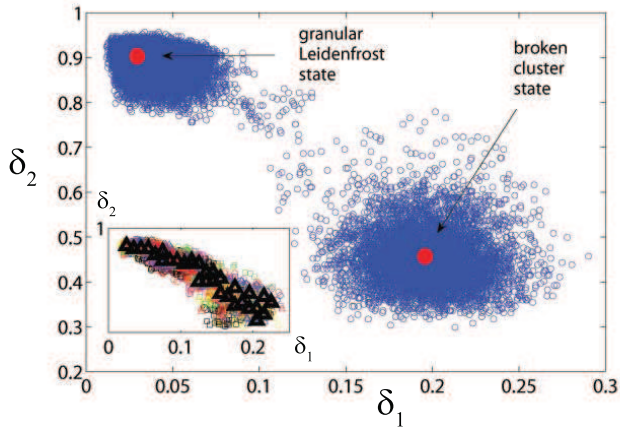}
\caption{System dynamics in the phase plane of effective reaction coordinates. The granular Leidenfrost state is in the upper left corner of the diagram ($\delta_1$ is close to $0$, $\delta_2$ is close to $1$), the broken cluster state is in the lower right corner. The dynamics of the system is characterized by a trajectory on this phase diagram; symbols represents the position of the system at a certain time (printed every $10 t_{MD}$). The system spends an exponentially long time near the first ``well" (unbroken cluster state), then quickly ``climbs the hill" and goes to the second ``well", spending an exponentially long time there. The inset shows $14$ trajectories corresponding to rare transitions between the two states.
\label{fig:cluster}
}
\end{center}
\end{figure}

There is an important difference between this description and the one-dimensional case presented in Figure 3.  In the  one-dimensional system, the reaction coordinate is exact. In our case there seems to be an effective two-dimensional system. The apparent fact that the reduced dynamics in this phase plane properly represents the true many-dimensional dynamics of the system is non-trivial. Put another way, there seems to be a more-or-less well defined transition path that is easy to demonstrate in these two coordinates.

Of course, we  do not know the position of the ``saddle point", i.e. of the separatrix between the basins of attractions of the two states. More than that, since we deal with the reduced dynamics, the saddle point might not be well defined and could occupy a region in the two-dimensional phase plane; this seems not to be the case. We claim that we have a sensible reaction coordinate since were able to compute the waiting time for the rare transition and verify this result in independent numerical simulations.

First, we tried to compute the average time for a transition from state $1$ to state $2$ directly, by simulating many systems for a very long time. We were able to see quite a few trajectories corresponding to these rare transitions, see the inset in Figure 5. However, despite the lengthy simulations, the transitions were observed in only $30$ percent of the systems, so the direct method was inefficient.

Then we implemented the ideas of FFS. Consider a single barrier, shown in Figure 6. This region (between the two straight lines in the phase plane) is ``very far" from state $1$ (the unbroken cluster state, top left corner), so the system visits this region very rarely. In order to compute the waiting time for the transition, we have to calculate two quantities. The first is how frequently the system visits this region in the phase plane. The second is the probability for a transition to state $2$ (the broken cluster state) provided that the system starts at the barrier (between the two straight lines) and does not go back to state $1$ before the transition occurs.

From direct simulations of $40$ systems, we found that systems crossed the barrier (uphill) $119$ times during the overall time of $T=126016530$, so the typical time to cross the barrier is $1.059 \times 10^6$. Next, we found $90$ different initial configurations inside the barrier (blue circles in Figure 6). We then chose $15$ representative systems (proportional to the density of points inside the barrier) and proliferated them. This means that instead of a single point (representing a single configuration of particles), we considered $10$ daughter systems, starting from the same configuration. Because particles collide stochastically with the bottom wall, these $10$ simulations rapidly depart from each other. Therefore, we have $10$ independent runs for each of the $15$ chosen configurations. In roughly $15$ percent of these $150$ runs, the system made a transition to the broken cluster state (before visiting the first ``well"); in other cases, the system was back to the granular Leidenfrost state. Based on these numbers, the overall waiting time for the rare transition between the granular Leidenfrost state and the broken cluster state is $1.0590\,10^6 / 0.15 = 7.0597\,10^6$.

\begin{figure}[ht]
\begin{center}
\includegraphics[width=3.5 in]{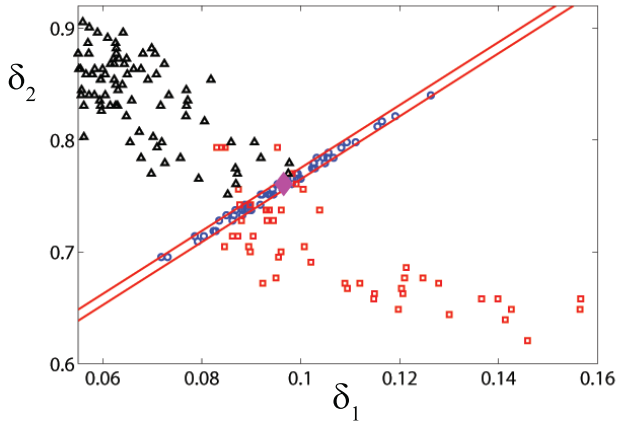}
\caption{Illustration of the FFS method. We show the barrier in the phase plane of two effective reaction coordinates. Blue circles represent different systems on the barrier. A representative fraction of these systems were proliferated, i.e. for each system we proliferated, we considered $10$ systems, starting from the same configuration. The magenta diamond represents one  such proliferated configuration. One of the daughter states (red squares) breaks,  while another daughter goes back to the granular Leidenfrost state (black triangles).
\label{fig:cluster}
}
\end{center}
\end{figure}

To check the validity of this method, we calculated the waiting time for the transition using a different approach, employed, for example, in computing distributions of escape events in swept-bias Josephson junctions \cite{method}. Consider an ensemble of N systems and consider time interval $\tau$, which is much larger than the system relaxation time, but much smaller than the waiting time for the transition. Running many simulations for time $\tau$, we can compute the fraction of runs, $f$, in which the transition did occur. Then the waiting time for the transition can be estimated as $\tau / f$. The result of approximately $10^7$ agrees well with the one obtained employing FFS with the reduced dynamics described by two effective reaction coordinates.

The method in the previous paragraph depends on finding a few successful crossing events. That is, it cannot be used too far from the transition. In contrast, the FFS-based method can give reliable results \emph{even when the transition times are so long that it is impractical to see even one crossing.} This is the virtue of the weighted-sampling method.

\section{Summary and discussion}
This work both advances the field of physics of granular matter and makes progress in the general understanding of rare events in  systems with a large number of degrees of freedom. First, we found that the granular Leidenfrost state can break somewhere in a spectacular volcanic-like explosion, while the remaining parts of a cluster do not melt and coexist with hot granular gas. A somewhat similar instability was observed in the classical Leidenfrost effect, where the droplet can not be too elongated in the horizontal direction, otherwise an instability occurs and the droplet splits into two parts; this phenomenon is sometimes called chimney instability \cite{chimney}.

Next, we found bistability in a wide range of external driving amplitudes: the two nonequilibrium steady states (the granular Leidenfrost state and the broken cluster state) exist for the same parameters in a wide range of parameter space. This gave us the possibility to investigate a rare event: a large fluctuation leading to cluster breaking, a sudden spontaneous transition from one state to another. We were able to find the reduced dynamics in terms of two effective reaction coordinates $\delta_1$ and $\delta_2$ and FFS with one barrier in the phase plane $(\delta_1,\delta_2)$ to compute the waiting time for this rare transition. This result was verified by using a different numerical method.

We also analyzed the dynamics of $y_{cm, i}$, the $y$ component of the center of mass for each vertical column. Measuring the power spectrum of these fluctuations, we found two low frequency peaks. The first peak corresponds to cluster oscillations, and the same peak occurs in the power spectrum of the total center of mass (these oscillations have recently been reported \cite{oscillations}). However, the second (much stronger) peak is not present in the power spectrum of the total center of mass, and is not related to oscillations of a cluster as a whole, but rather to the wave-like modes along the cluster. These modes are stable below the instability threshold, but are constantly driven by stochastic noise. One can speculate that the rare cluster breaking might be related to an unusually large energy input to these modes (rare large energy fluctuation).

\begin{acknowledgments}
E.K. thanks Misha Khasin and Oleg Kogan for fruitful discussions.
\end{acknowledgments}


\end{document}